\documentclass[aps,prb,showpacs,reprint]{revtex4-1}

\usepackage{graphicx}
\usepackage{amsmath}
\usepackage{bbm}

\begin{document}
\title
{Spin transistor operation driven by the Rashba spin-orbit
coupling in the gated nanowire}
\author{P. W\'ojcik}
\author{J. Adamowski}
\email[Electronic address: ]{adamowski@fis.agh.edu.pl}
\author{B.J. Spisak}
\author{M. Wo{\l}oszyn}
\affiliation{AGH University of Science and Technology, Faculty of
Physics and Applied Computer Science, al. Mickiewicza 30,
Krak\'ow, Poland}

\begin{abstract}
The theoretical description has been proposed for the operation
of the spin transistor in the gate-controlled InAs nanowire.
The calculated current-voltage characteristics show that
the current flowing from the source (spin injector) to the drain
(spin detector) oscillates as a function of the gate voltage, which
results from the precession of the electron spin caused by the
Rashba spin-orbit interaction in the vicinity of the gate.
We have studied two operation modes of the spin transistor: (A) the ideal operation mode with the full
spin polarization of electrons in the contacts, the zero temperature, and the single
conduction channel corresponding to the lowest-energy subband of
the transverse motion and (B) the more realistic operation mode
with the partial spin polarization of the electrons in the
contacts, the room temperature, and the conduction via many transverse subbands taken
into account.
For mode (A) the spin-polarized current can be switched on/off by the suitable
tuning of the gate voltage, for mode (B) the current also exhibits
the pronounced oscillations but with no-zero minimal values.
The computational results obtained for mode (B) have been compared with the recent experimental data
and a good agreement has been found.
\end{abstract}

\maketitle
\section{Introduction}

The all-electrical control of the spin-polarized current is a
basic principle of the operation  of novel spintronic devices
including the spin transistor proposed by Datta and
Das.\cite{Datta90} According to this idea~\cite{Datta90} the
current of spin polarized electrons injected from the
ferromagnetic source contact into a semiconductor is modulated by
the Rashba spin-orbit interaction (SOI)\cite{Rashba84} in a
conduction channel. The state of the spin transistor depends on
the spin orientation of the electron in the conduction channel
with respect to the magnetization of the ferromagnetic drain
contact. The successful operation of the spin transistor requires
an efficient spin injection from the ferromagnetic
source\cite{Matsu02} into the semiconductor and an effective
method of manipulation of the electron spin in the conduction
channel. The efficiency of the spin-polarized current injection
from the ferromagnetic electrode into the semiconductor is rather
low.\cite{Schmidt} In order to improve this efficiency the
semiconductor spin filter based on the dilute magnetic
semiconductor heterostructure  has been applied.\cite{Slobod03}
The operation of the paramagnetic resonant tunneling diode (RTD)
as the spin filter at low temperature was demonstrated
experimentally by Slobodskyy et al.\cite{Slobod03} and
investigated theoretically in our previous paper.\cite{Wojcik12}
However, the operation of the  paramagnetic RTD as the spin filter
requires a low temperature and a high magnetic
field.\cite{Slobod03,Wojcik12} The fabrication of the spin filter
from the ferromagnetic semiconductor allows to extend the
applicability of the spin filter operation to the room temperature
without the external magnetic field.  In our recent
paper,\cite{Wojcik13} we have studied the spin filter effect in
the mesa-type RTD based on ferromagnetic GaMnN and shown that --
in this device -- the spin filter operation can be realized at the
room temperature.

The effective manipulation of the electron spin in a semiconductor
can be realized by the Rashba SOI that couples the linear momentum
of the electron with its spin via the external electric field. In
the last years, the application of the SOI to the manipulation of
electron spins in semiconductor nanodevices has been a subject of
numerous
papers.\cite{Schlie03,Flindt06,Liu06,Fasth07,Liu07,Ohno07,Ohno08,Nazmit09,Gelabert11,Bringer11,Thorg12,Yoh12,Ban13,Sadreev13}
In Ref.~\onlinecite{Gelabert11}, the conductance oscillations due
to the Rashba SOI have been investigated in two-dimensional
stripes. The electron spin modification in semiconductor nanowires
has been studied in Refs.\cite{Bringer11,Ban13,Sadreev13} Bringer
and Sch\"{a}pers\cite{Bringer11} demonstrated the spin precession
and modulation in a cylindrical nanowire resulting from the Rashba
SOI. The shape-dependent spin transport has been studied in
semiconductor nanowires without and with the SOI taken into
account.\cite{Ban13} Moreover, the joint effect of the ac gate
voltage, constant magnetic field, and the Rashba SOI on the
electron transmission in the nanowire has been considered in
Ref.\onlinecite{Sadreev13}. These studies have led to a
realization of the spin transistor in the semiconductor layer
heterostructures\cite{Wund10,Bett12} and nanowires.\cite{Yoh12}
Recently, Yoh et al.\cite{Yoh12} have fabricated the spin
transistor based on the InAs nanowire and reported the
gate-voltage controlled on/off switching of the current.

In this paper, we present the computational results for the spin transistor operation
in InAs nanowire with the side gate electrode.
These results allow us to extract the most important physical properties that underly the operation
of the nanowire spin transistor.
We show that the gate-voltage induced SOI in the nanowire leads to the
current oscillations as a function of the gate voltage.
We have studied the spin transistor operation
under the ideal conditions, i.e., the full spin polarization of the electrons in the contacts
at zero temperature using the one-subband approximation,
and the more realistic conditions with the partial spin electron polarization in the contacts
at a room temperature in the many-subband approximation.
The results obtained for the second case have been compared with the experimental data.
The paper is organized as follows:  in Section II, we present the theoretical model,
the results are presented in Section III, Section IV contains the conclusions and summary.

\section{Theory}

\begin{figure}[ht]
\begin{center}
\includegraphics[scale=0.3, angle=-90]{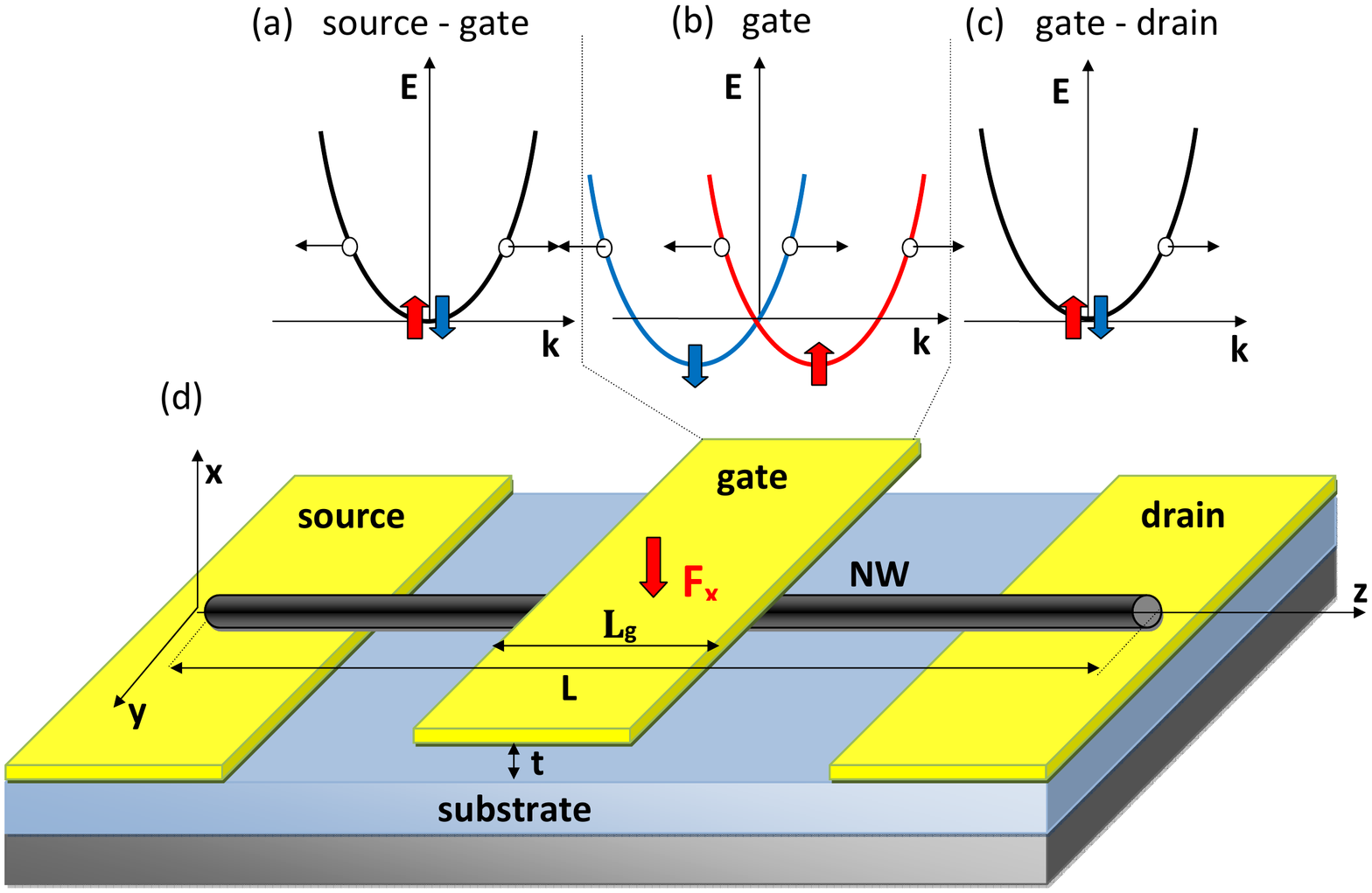}
\caption{Dispersion relations $E(k)$ for (a) source-gate, (b) gate, and (c) gate-drain
regions of the nanowire. In panels (a-c), the vertical (red and blue) arrows show the spin orientations, while the horizontal arrows
show the direction of the electron motion.  (d) Schematic of the nanowire spin transistor.
Electric field $F_x$ of the gate acts in the central (gate) region of the nanowire with length $L_g$.
$L$ is the length of the nanowire.}
\label{fig1}
\end{center}
\end{figure}

We consider the InAs nanowire with the nearby gate
that generates the transverse homogeneous electric field
of finite range (Fig.~\ref{fig1}).
The  one-electron Hamiltonian has the form
\begin{equation}
H=-\frac{\hbar ^2}{2m_e} \nabla ^2 + U(\mathbf{r}) + H_R(\mathbf{r}) \;,
\label{H}
\end{equation}
where $m_e$ is the effective mass of the conduction band electron and $\mathbf{r} = (x,y,z)$.
The potential energy of the electron
\begin{equation}
U(\mathbf{r}) = U_c(x,y) + U_{ds}(z) + U_g(x,z)
\label{U}
\end{equation}
is the sum of the lateral (transverse) confinement potential energy $U_c(x,y)$,
potential energy $U_{ds}(z) = e F_z z$ in the (longitudinal) electric field
$F_z = - V_{ds} /L$, where $V_{ds}$ is the drain-source voltage,
$e$ is the elementary charge ($e>0$), and $L$ is the length of the nanowire, and
potential energy $U_g(x,z)= e g(z) F_x x$ in the (transverse) electric field
$F_x$ generated by the gate, where function $g(z)$ determines the potential energy profile
along the nanowire axis.
In the present paper, we take on
\begin{equation}
g(z) = \exp\left[-\left(\frac{|z-L/2|}{L_g/2}\right)^p\right] \;.
\label{g}
\end{equation}
For sufficiently large $p$ ($p>10$) $g(z) \simeq 1$ in the central region
of the nanowire, i.e.,  near the gate of width $L_g$ [cf. Fig.~\ref{fig1}(d)],
and $g(z) \simeq 0$ otherwise.
The last term in Eq.~(\ref{H}) describes the Rashba SOI and has the form
\begin{equation}
H_R(\mathbf{r}) = \gamma \nabla U(\mathbf{r}) \cdot (\mbox{\boldmath ${\hat{\sigma}}$} \times \mathbf{\hat{k}}) \;,
\label{HR}
\end{equation}
where $\gamma$ is the Rashba coupling constant,
$\mbox{\boldmath ${\hat{\sigma}}$}= (\hat{\sigma}_x, \hat{\sigma}_y, \hat{\sigma}_z)$ is the Pauli matrix vector,
and $\mathbf{\hat{k}} = -i \nabla = (\hat{k}_x, \hat{k}_y, \hat{k}_z)$.
We assume that the lateral confinement potential is strong and has the hard-wall shape.
This means that $U_c(x,y)$ is flat inside the nanowire, i.e., in Eq.~(\ref{HR}), the first derivative
of $U_c(x,y)$ disappears inside the nanowire,
while the wave function of the transverse size-quantized
motion vanishes at the surface of the nanowire.  Therefore,
we neglect the contribution of lateral confinement to the Rashba Hamiltonian (\ref{HR}). Moreover,
the contribution stemming from the longitudinal drain-source electric field $F_z$
is also negligible, since this field is a few orders of magnitude smaller
than the electric field $F_x$ generated by the gate. \\
Under these assumptions Hamiltonian (\ref{HR}) takes on the form
\begin{eqnarray}
 H_R&=&\gamma e F_xg(z)(\hat{\sigma}_y \hat{k}_z - \hat{\sigma}_z \hat{k}_y) \nonumber \\
&+&\gamma e F_x x \frac{dg(z)}{dz} (\hat{\sigma}_x \hat{k}_y-\hat{\sigma}_y \hat{k}_x).
\label{HR2}
\end{eqnarray}
We calculate the expectation value of $H$ in the transverse state $m$ and obtain
\begin{equation}
 H_{R,1D} = -i  e \gamma F_x g(z) \hat{\sigma}_y \frac{d}{dz}-\frac{i\gamma e F_x \hat{\sigma}_y}{2} \frac{dg(z)}{dz} \;,
 \label{R1D}
\end{equation}
where the last term results from the position-dependent SOI and assures  the hermiticity of the Hamiltonian.
Finally, the effective one-dimensional Hamiltonian takes on the form
\begin{eqnarray}
H_{e\!f\!f} &=&-\frac{\hbar ^2}{2m_e} \frac{d^2}{dz^2}+ U_{ds}(z) + eg(z)F_xt+E^{\perp}_m  \nonumber \\
&-& i  e \gamma F_x g(z) \hat{\sigma}_y \frac{d}{dz}-\frac{i\gamma e F_x \hat{\sigma}_y}{2} \frac{dg(z)}{dz} \;,
\label{H1D}
\end{eqnarray}
where $E^{\perp}_m$ is the energy of the $m$-th transverse state and $t$ is the average distance between the gate electrode and nanowire

We have calculated the spin-dependent reflection and transmission coefficients for subband $m$ using the transfer matrix method
on the grid with mesh points $z_n = n L/(\mathcal{N}-1)$, where $n = 0, \ldots , \mathcal{N}$ and $\mathcal{N}$ has been taken to be $10^4$.
According to the transfer matrix method, we assume that the wave function and the current density are continuous
at the interfaces between subintervals $(z_{n-1},z_{n})$ and $(z_{n},z_{n+1})$.
The continuity condition for the wave function $\Psi^m(z)$ corresponding to the transverse state $m$
has the form
\begin{equation}
\Psi^m_n(z=z_n)= \Psi^m_{n+1}(z=z_n) \; ,
\label{cond1}
\end{equation}
where $\Psi^m_n(z)$ is the $m$-subband  wave function in the subinterval $(z_{n-1},z_{n})$.
The continuity of the current density is equivalent to the following condition:
\begin{equation}
\hat{v} \Psi^m_n\big|_{z=z_n} = \hat{v}\Psi^m_{n+1} \big|_{z=z_{n}} \;,
\label{cond2}
\end{equation}
where $\hat{v}$ is the velocity operator, which in the presence of the SOI has the form
\begin{equation}
\hat{v}=\frac{\hbar}{im_e}\frac{d}{dz}+\frac{e \gamma F_x}{\hbar}\hat{\sigma}_y \; .
\end{equation}
When applying conditions (\ref{cond1}) and (\ref{cond2}) we assume that the SOI is constant
in each subinterval $(z_{n-1}, z_{n})$, which is compatible with the basic assumption of the transfer matrix method.
In each subinterval $(z_{n-1}, z_{n})$, the wave function $\Psi^m_n(z)$ has the spinor form
\begin{eqnarray}
\Psi^m_n(z)=  a_n \left (
\begin{array}{c}
u_n^+\\
v_n^+
\end{array}
\right )
e^{ik_n^+z}
 + b_n
\left (
\begin{array}{c}
u_n^+\\
v_n^+
\end{array}
\right )
e^{-ik_n^+z} \nonumber  \\
+ c_n \left(
\begin{array}{c}
u_n^-\\
v_n^-
\end{array}
\right)
e^{ik_n^-z}
+ d_n
\left (
\begin{array}{c}
u_n^-\\
v_n^-
\end{array}
\right )
e^{-ik_n^-z} \;,
\label{Psi}
\end{eqnarray}
where the $z$ component of the wave vector  is given by
\begin{equation}
k_n^{\pm} = \frac{m_e}{\hbar ^2} \sqrt{2 \left[ \frac{\hbar ^2(E-U_n)}{m_e}
+  e^2 \gamma^2 F^2_x g_n^2  \pm \sqrt{\Delta_n} \right]}
\end{equation}
with
\begin{equation}
\Delta_n = e^2 \gamma^2 F^2_x g_n^2 \left[ \frac{2\hbar ^2(E-U_n)}{m_e}
+ e^2 \gamma^2 F^2_x g_n^2 \right] \;,
\end{equation}
where $E$ is the energy of the electron, $U_n= U_{ds}(z=z_n)+E^{\perp}_m+eg_nF_xt$, and $g_n=g(z=z_n)$.
The dispersion relations $E(k)$ for the different regions of the nanowire
are depicted in Fig.~\ref{fig1}(a-c).
The quantum states of the electrons with opposite spins are degenerate in
the source-gate and gate-drain regions.  This degeneracy is lifted
near the gate due to the gate-induced spin-orbit interaction.\\
The spin up ($+$) and spin down ($-$) amplitudes are given by
\begin{eqnarray}
u_n^{\pm} &=& \frac{ \left [ \frac{\hbar ^2 (k_n^{\pm})^2}{2m_e}
+
(E-U_n) \right ]^2}{ \left [\frac{\hbar ^2 (k_n^{\pm})^2}{2m_e} - (E-U_n) \right ]^2 + (k_n^{\pm} eF_x g_n)^2} \;, \\
v_n^{\pm} &=& \frac{[k_n^{\pm} e F_x g_n]^2}{ \left [\frac{\hbar ^2 (k_n^{\pm})^2}{2m_e} - (E-U_n) \right ]^2
+ (k_n^{\pm} e F_x g_n)^2} \;.
\end{eqnarray}
For each subband $m$ we calculate the probabilities of the following processes:
the reflection with the spin conservation, i.e., no-spin-flip reflection ($R^m_{\uparrow \uparrow}$ and $R^m_{\downarrow \downarrow}$),
the reflection with the spin flip  ($R^m_{\uparrow \downarrow}$  and $R^m_{\downarrow \uparrow}$),
the transmission with the spin conservation, i.e., no-spin-flip transmission ($T^m_{\uparrow \uparrow}$  and $T^m_{\downarrow \downarrow})$,
and the transmission with the spin flip  ($T^m_{\uparrow \downarrow}$  and $T^m_{\downarrow \uparrow}$).
Here and in the following part of the paper, we are using subscripts $\sigma = \uparrow$ ($\downarrow$)
for the spin up (down) states.
In these calculations, we have neglected the intersubband transitions.\\

Having determined the transmission coefficients $T^m_{\sigma\sigma^{\prime}}(E)$
we have calculated the current in the ballistic regime using the Landauer formula
\begin{equation}
 I_{\sigma \sigma^{\prime}}= \frac{2e}{h} \sum\limits_{m=1}^M \int\limits_0^{\infty} dE
 \:T^m_{\sigma \sigma^{\prime}}(E) [f_s(E, \mu_s)-f_d(E,\mu_d)] \;,
\label{I}
\end{equation}
where $\sigma$ and $\sigma^{\prime}$ corresponds to the spin of the electron
in the source and drain, respectively,
and $M$ is the maximum number of the transverse subbands taken into account.
In Eq.~(\ref{I}), $f_{s,d}(E, \mu_{s,d})$ is the Fermi-Dirac distribution function for the electrons in the source ($s$)
with chemical potential $\mu_s$ and drain ($d$) with chemical potential $\mu_d$.
In order to describe the polarizing effect of the
contacts we introduce polarization $P$ defined as follows:
\begin{equation}
P = \frac{n_{\uparrow}-n_{\downarrow}}{n_{\uparrow}+n_{\downarrow}} \;,
\end{equation}
where $n_{\sigma}$ is the concentration of electrons with spin $\sigma$ at the Fermi level.
In the present paper, we neglect the resistance at the ferromagnet/semiconductor interface,
which means that the electrons with the well-defined spin are injected from the
ferromagnetic source into the nanowire with the 100\% efficiency.
Therefore, the spin current components $I_{\sigma}$ are calculated as follows:
\begin{equation}
I_{\uparrow} = \frac{1+P}{2} I_{\uparrow\uparrow} + \frac{1-P}{2} I_{\downarrow\uparrow}
\label{I_up}
\end{equation}
and
\begin{equation}
I_{\downarrow} = \frac{1+P}{2} I_{\uparrow\downarrow} + \frac{1-P}{2} I_{\downarrow\downarrow} \;.
\label{I_down}
\end{equation}
The total current is given by
\begin{equation}
I_{total}= \frac{1+P}{2} I_{\uparrow} + \frac{1-P}{2} I_{\downarrow} \;.
\label{I_total}
\end{equation}

The present calculations have been performed for the InAs nanowire with the following material parameters:
$m_e=0.026m_0$, where $m_0$ is the free electron mass, $\gamma=1.17$~nm$^2$, and $\mu_s = \mu_d = \mu$.
In order to fulfil the conditions of the ballistic transport,
we have chosen the geometric parameters of the nanowire as follows:
$L=250$~nm and $L_g = 150$~nm.
The transverse states have been calculated under assumption that the nanowire
possesses the square-shape cross-section with side width $d=40$~nm.
Moreover, the infinite hard-wall potential in the transverse direction has been assumed.
Since the gate can be treated as the plane capacitor with interelectrode distance $t=30$~nm,\cite{Yoh12} we convert electric field $F_x$
of the gate to gate voltage $V_g$ as follows: $V_g = - F_x t$, which allows us to present the results as functions
of the gate voltage.

\section{Results}

In this section, we present the results for the two operation modes of the spin transistor.
In Subsection A, we present the results for the ideal operation mode, i.e., for
the full spin polarization ($P=1$) of the electrons at the Fermi level in the source and drain contacts.
We assume the zero temperature and the single-channel electron transport, i.e., the conduction  via
the lowest-energy subband of the transverse motion.  In Subsection B, we present the computational results
for the more realistic operation mode, i.e., for the
partial  polarization ($P=0.4$) of electron spins in the contacts
and compare them with the experimental data.\cite{Yoh12}
These calculations have been performed at the room temperature in the framework of the many subband approximation.

\subsection{Ideal operation mode}

The one-subband approximation takes into account only the
lowest-energy subband of the transverse size-quantized motion.
This approximation can be justified for the strong lateral
confinement that occurs in very thin nanowires, in which the
electron occupies the lowest-energy transverse quantum state. In
these calculations, we assume that only the electrons
with spin up are injected from the source into the nanowire and
can be detected in the drain. Moreover, we take on $m=0$ (for
simplicity we omit index $m$) and choose $\mu = 10$ meV.

\begin{figure}[ht]
\begin{center}
\includegraphics[scale=0.25, angle=0]{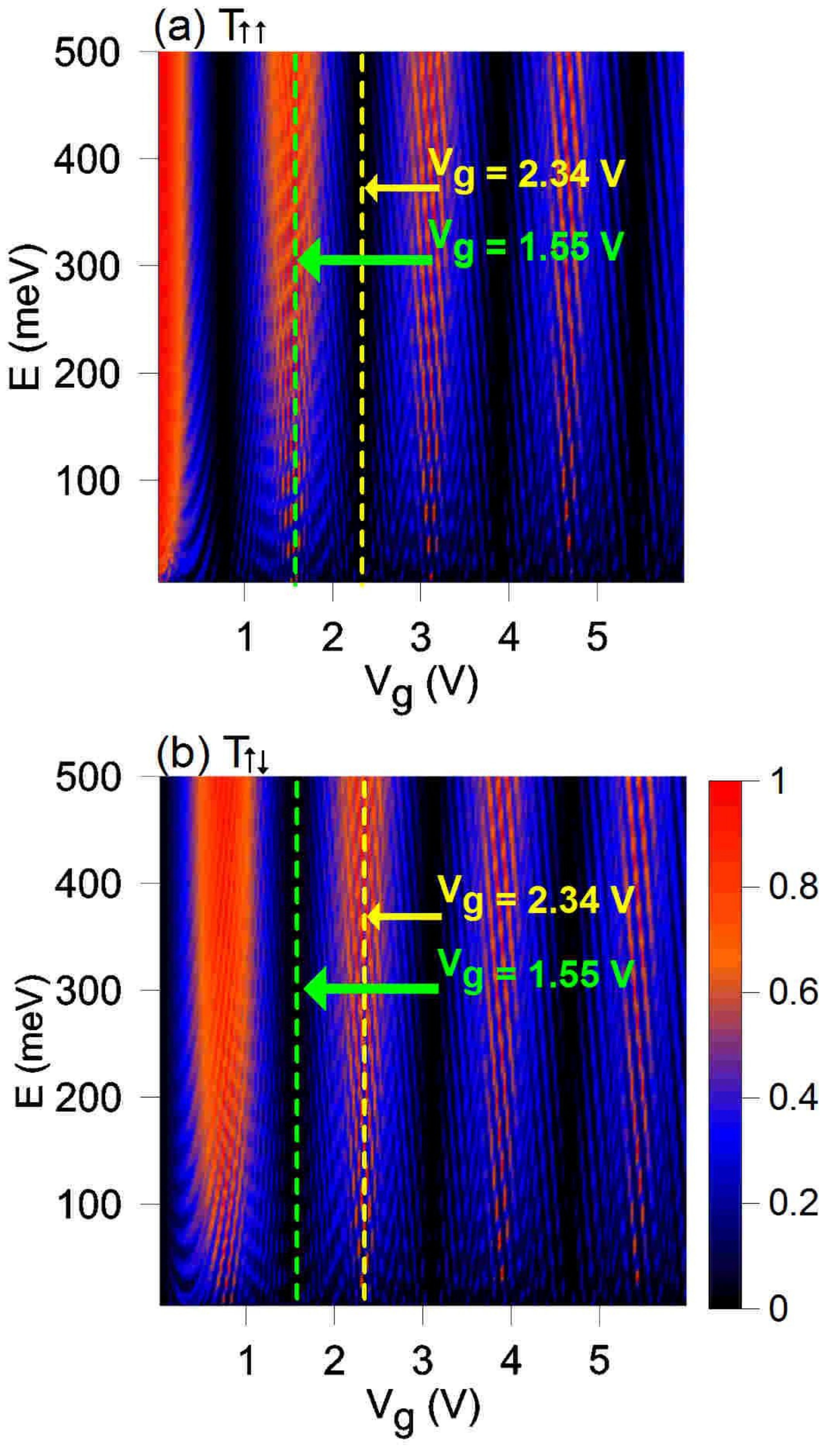}
\caption{Transmission coefficients for the transitions (a) with no-spin-flip $T_{\uparrow \uparrow}$
and (b) spin-flip $T_{\uparrow \downarrow}$
as functions of incident electron energy $E$ and gate voltage $V_g$.
The vertical broken lines correspond to the  maxima
and minima of the transmission coefficients.}
\label{fig2}
\end{center}
\end{figure}

\begin{figure}[ht]
\begin{center}
\includegraphics[scale=0.25, angle=0]{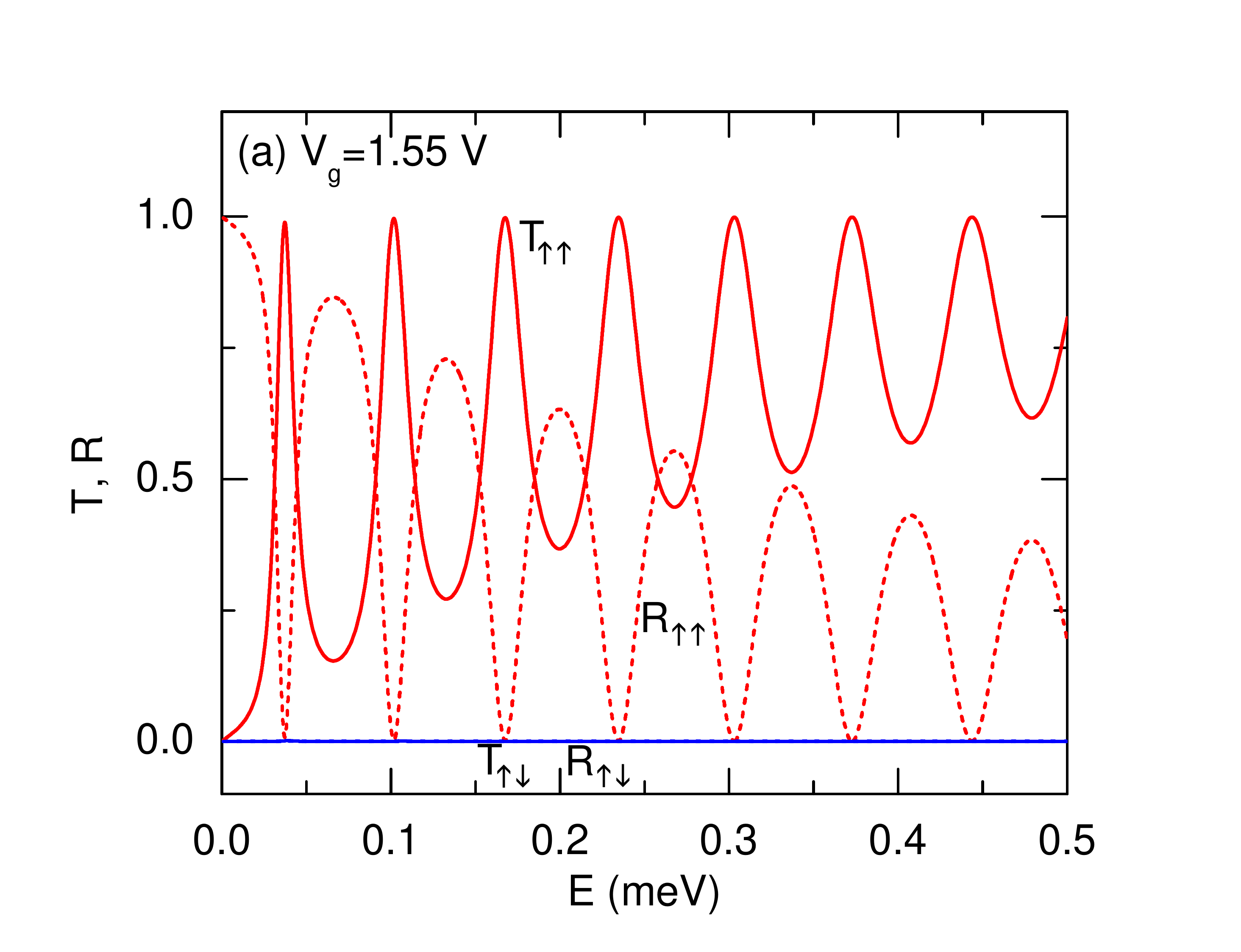}
\includegraphics[scale=0.25, angle=0]{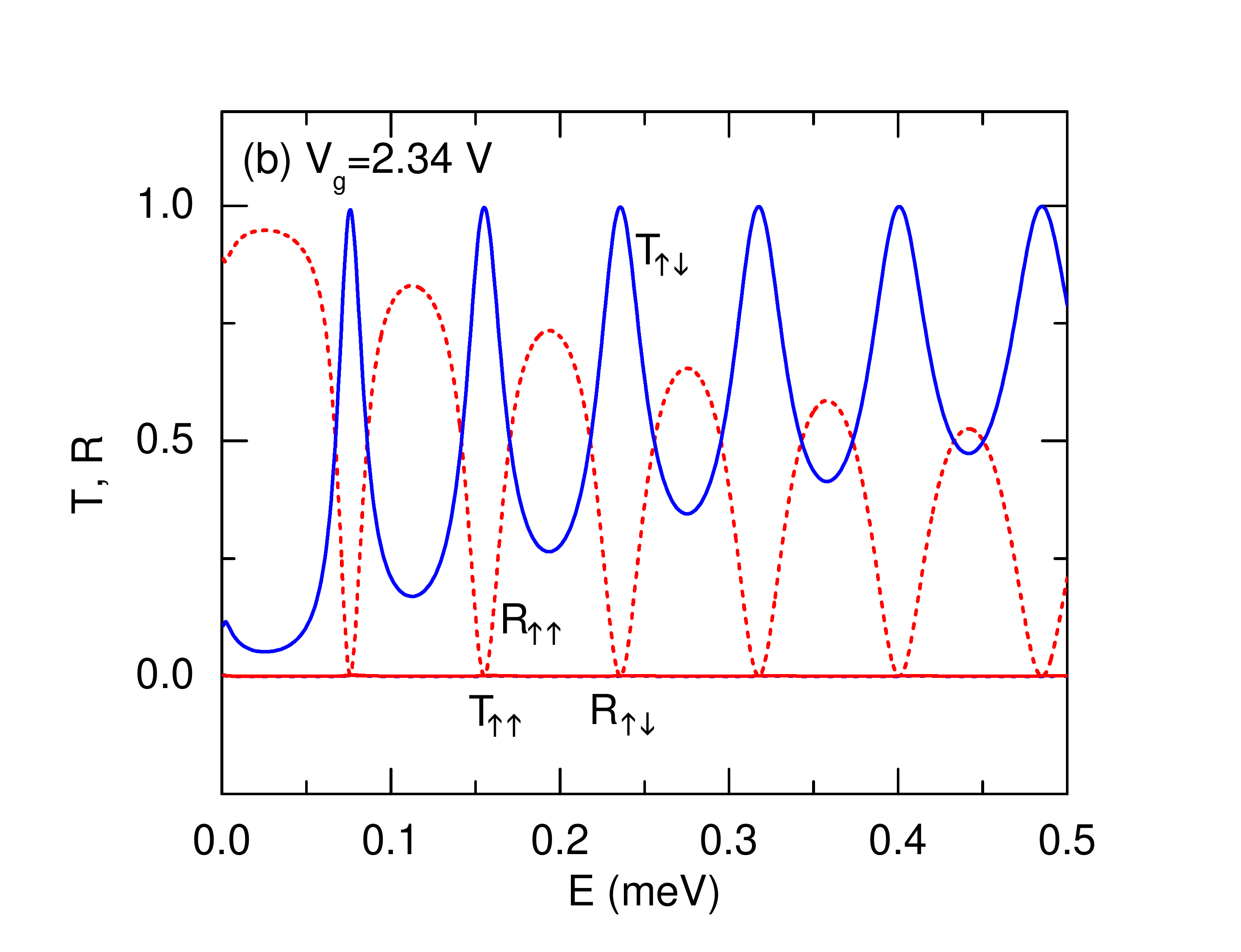}
\caption{Transmission $T$ and reflection $R$ coefficients
as functions of energy $E$ of the injected electron for (a) $V_g = 1.55$~V
and (b) $V_g = 2.34$~V.}
\label{fig3}
\end{center}
\end{figure}

In Fig.~\ref{fig2} we present the transmission coefficients for the
no-spin-flip [Fig.~\ref{fig2}(a)] and spin-flip [Fig.~\ref{fig2}(b)]
transitions.
Fig.~\ref{fig2} shows that the transmission coefficients $T_{\uparrow \uparrow}$
and $T_{\uparrow \downarrow}$ are oscillating functions of the gate voltage
with the period of the oscillations $\Delta V_g \simeq 1.6$ V.
The spin-flip and no-spin-flip transmission coefficients
oscillate in anti-phase.
We have chosen the values of the gate voltage that correspond to the maximum ($V_g=1.55$~V) and minimum ($V_g=2.34$~V)
of the transmission coefficient $T_{\uparrow \uparrow}$ (cf. vertical dashed lines in Fig. \ref{fig2}).
The corresponding transmission and reflection coefficients as a function of
the incident electron energy are depicted in Fig.~\ref{fig3},
which shows that the transmission and reflection coefficients
are oscillating functions of the incident electron energy.
The amplitude of these oscillations decreases with the increasing energy.
For $V_g = 1.55$~V the no-spin-flip transmission $T_{\uparrow \uparrow}$ and reflection $R_{\uparrow \uparrow}$ coefficients
oscillate taking on the values between 0 and 1, while the corresponding spin-flip
coefficients ($T_{\uparrow \downarrow}$ and $R_{\uparrow \downarrow}$) are equal to zero
[Fig.~\ref{fig3}(a)]. The opposite behaviour is observed in Fig.~\ref{fig3}(b),
in which the spin-flip transport dominates and the spin-conserved transport
vanishes.
\begin{figure}[ht]
\begin{center}
\includegraphics[scale=0.25, angle=0]{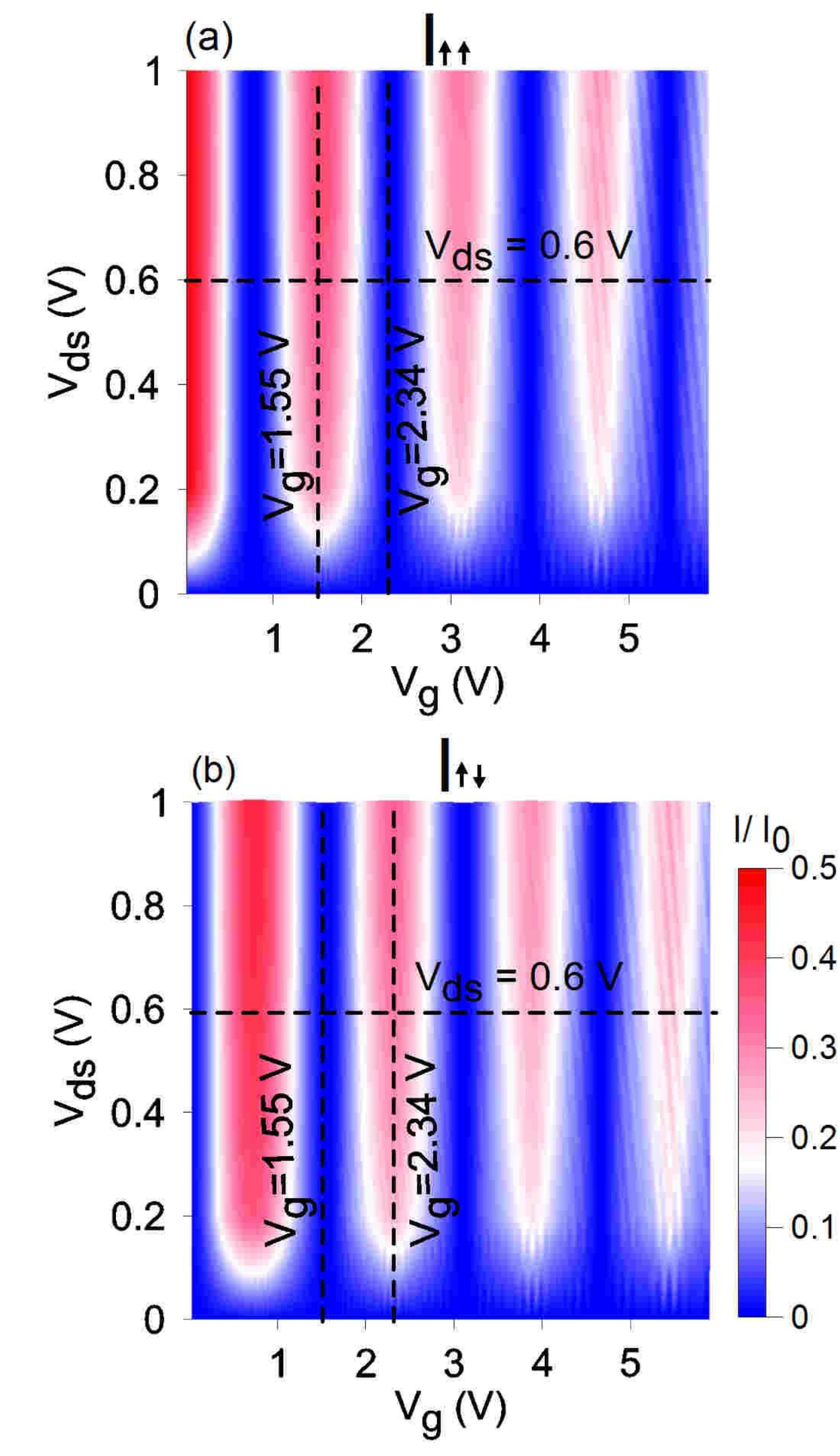}
\caption{Spin polarized current components as functions of the gate voltage $V_g$
and the drain-source voltage $V_{ds}$. The results for (a) spin-conserved current component $I_{\uparrow\uparrow}$,
(b) spin-flip current component $I_{\uparrow\downarrow}$.}
\label{fig4}
\end{center}
\end{figure}

\begin{figure}[ht]
\begin{center}
\includegraphics[scale=0.25, angle=0]{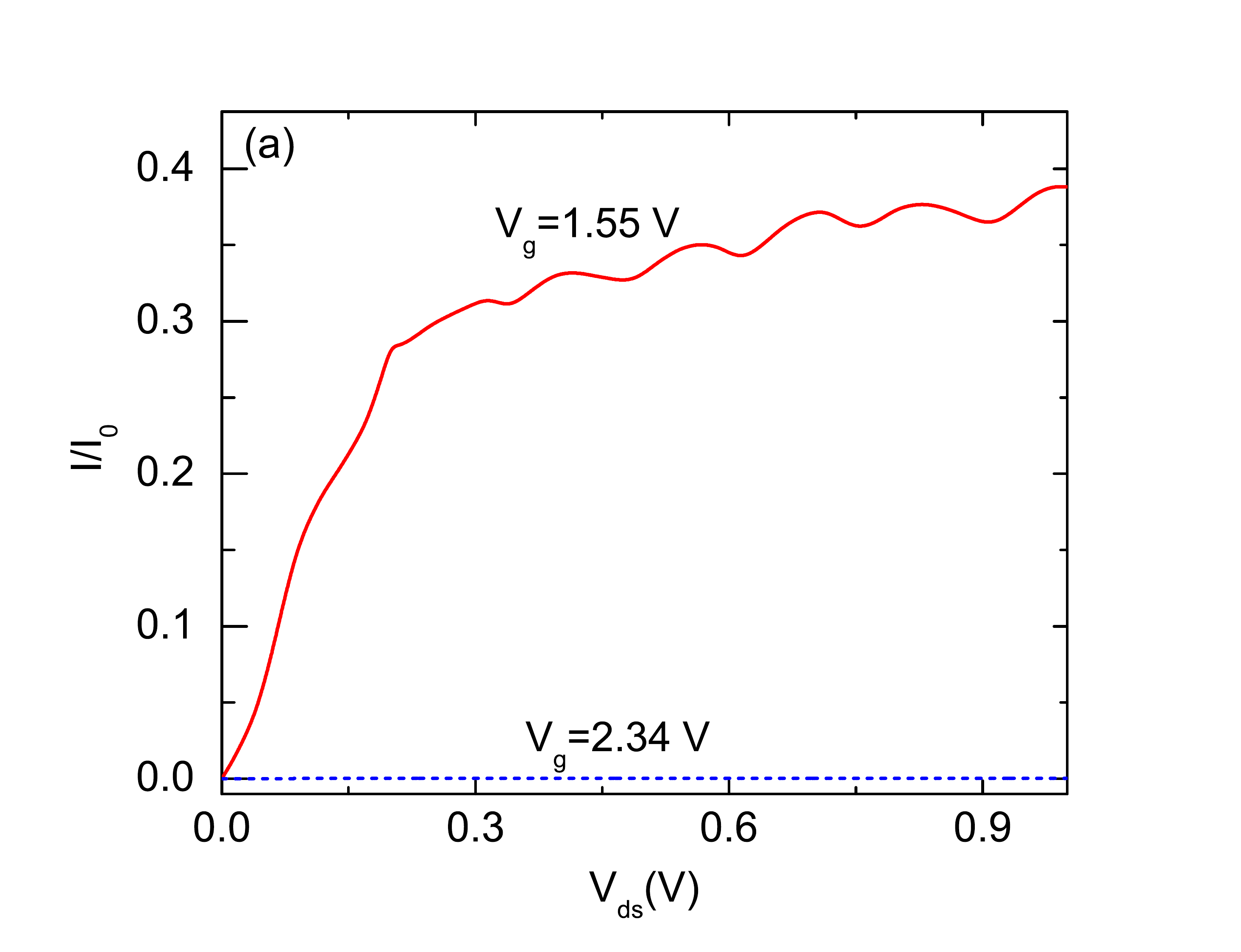}
\includegraphics[scale=0.25, angle=0]{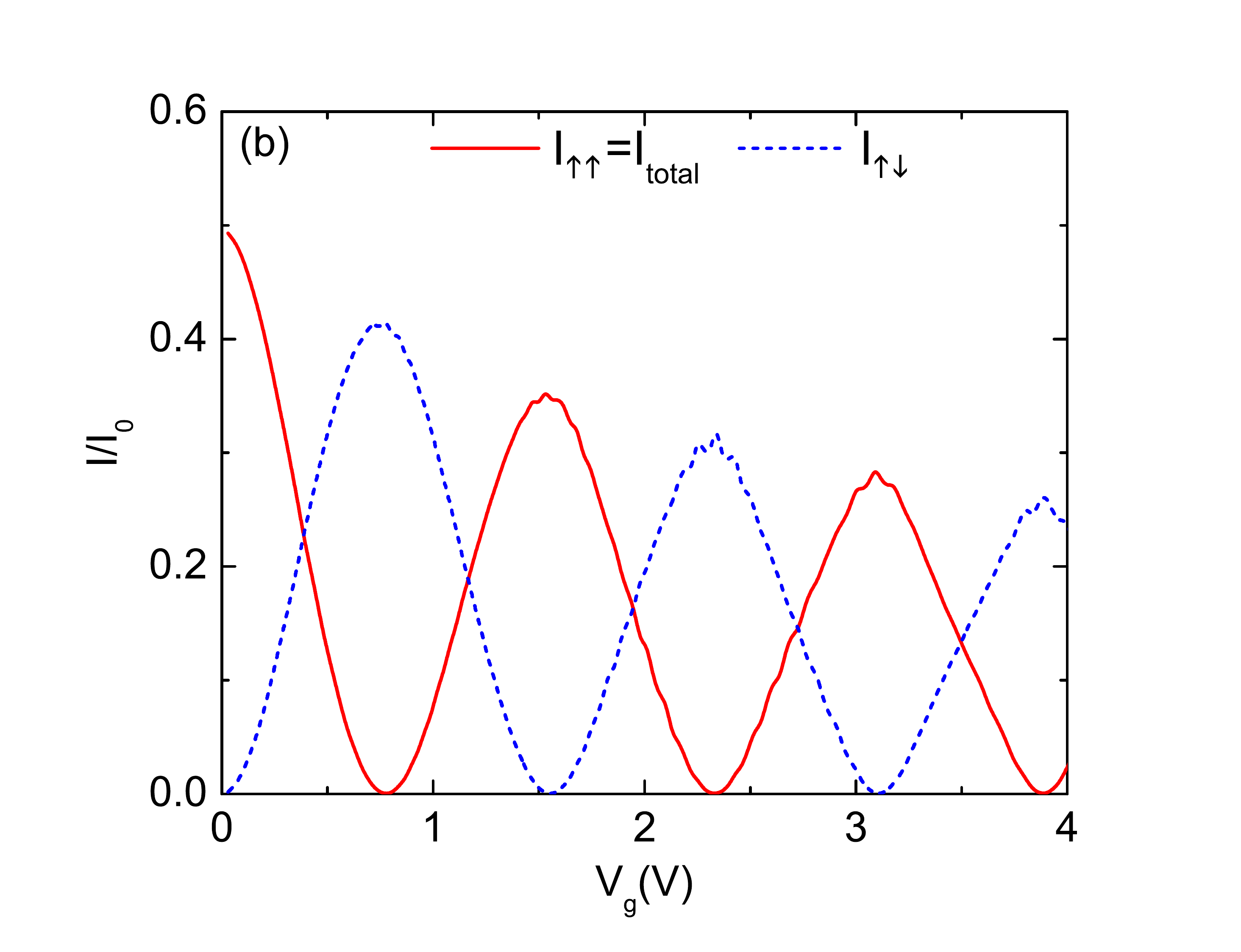}
\caption{(a) Current-voltage characteristics for $V_g = 1.55$~V and $V_g = 2.34$~V.
$I_{total} = I_{\uparrow\uparrow}$,
$V_{ds}$ is the drain-source voltage, $I_0 = 2e\mu/h$.
The plots are drawn along the vertical dashed lines on Fig.~\ref{fig4}(a).
(b) Spin-conserved ($I_{\uparrow\uparrow}$) and spin-flip ($I_{\uparrow\downarrow}$) current
as a function of the gate voltage $V_g$.  The plots are drawn along the horizontal lines $V_{ds}= 0.6$~V
on Figs.~\ref{fig4}(a) and \ref{fig4}(b).}
\label{fig5}
\end{center}
\end{figure}

The oscillations of the transmission presented in Fig.~\ref{fig2}
lead to the similar oscillations of the spin-polarized currents as
functions of the gate voltage.  Figure~\ref{fig4} displays spin-polarized
current components $I_{\uparrow\uparrow}$ and
$I_{\uparrow\downarrow}$ as functions of the drain-source voltage
and the gate voltage. We see that spin-flip
$I_{\uparrow\downarrow}$ and no-spin-flip $I_{\uparrow\uparrow}$
currents oscillate in anti-phase with the minimal values equal zero.
The current-voltage characteristics calculated for the chosen values of gate voltage
are plotted in Fig.~\ref{fig5}(a).
We note that only the spin-up electrons can pass through the drain,
i.e., total current $I_{total} = I_{\uparrow\uparrow}$, which results from
Eqs.~(\ref{I_up}), (\ref{I_down}), and (\ref{I_total}) for $P=1$.
The curve of $I_{total} = I_{\uparrow\uparrow}$ drawn for $V_g = 1.55$~V corresponds to the on-state of the spin transistor
and exhibits the typical features for the transistor operation:
for the low drain-source voltage the current rapidly increases
and becomes saturated for the higher drain-source voltage.
The small oscillations of the current for $V_{ds} \geq 0.3$~V result from the transmission via
the higher-energy resonance states formed in the quantum-well region
near the gate.
For $V_g = 2.34$~V, $I_{total} = I_{\uparrow\uparrow}=0$, which means that the spin transistor
is in the off-state.
Figure \ref{fig5}(b) shows that total  current $I_{total} = I_{\uparrow\uparrow}$
is the oscillating function of the gate voltage.
The spin-flip current oscillates in anti-phase and -- due to the polarizing effect of the drain --
can not flow in the circuit.
The results of Fig.~\ref{fig5}(b) demonstrate that the spin-polarized current can be switched on/off
by the suitable tuning of the gate voltage.\\

\begin{figure}[ht]
\begin{center}
\includegraphics[scale=0.3, angle=0]{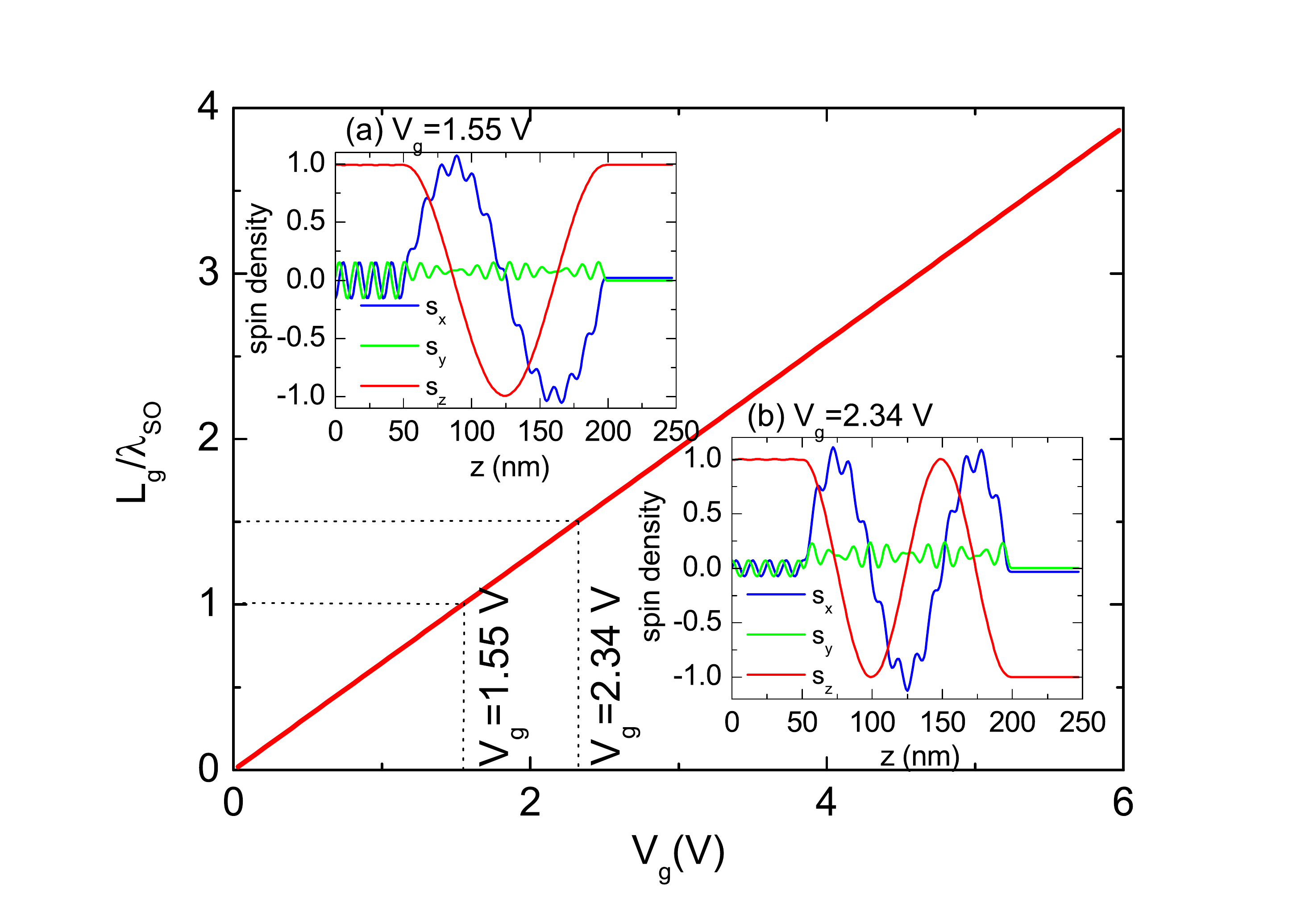}
\caption{Ratio $L_g/\lambda_{SO}$ as a function of the gate voltage $V_g$.
Solid (red) line is parametrized by the linear function
$L_g/\lambda_{SO} = \alpha V_g$ with $\alpha = 0.65$~V$^{-1}$.
Insets: Spin density components $s_x$ (blue curve), $s_y$ (green curve) , and $s_z$ (red curve)
as functions of coordinate $z$ measured along the nanowire axis for (a) $V_g = 1.55$V
and (b) $V_g = 2.34$V. The spin of the electron injected from the source is equal
to $s_z=+\hbar/2$.}
\label{fig6}
\end{center}
\end{figure}

In order to get a more physical insight into the electron spin dynamics in the nanowire
we have calculated the spin density, which is defined as follows:
\begin{equation}
s_j = \Psi^{\dagger} \hat{s}_j \Psi \;,
\label{spin_dens}
\end{equation}
where $\hat{s}_j = (\hbar/2) \hat{\sigma}_j$ ($j=x,y,z$)
and $\Psi^{\dagger}$ is the Hermitian conjugate of spinor $\Psi$.
The results presented in Fig.~\ref{fig6} [insets (a) and (b)]
show that the spin precession occurs in the gate region,
i.e., in the interval 50~nm $< z <$ 200~nm.
In both the cases [Fig.~\ref{fig6}(a,b)], the electron injected with spin $s_z=+\hbar/2$
changes its spin via the changes of the $s_x$ component,
while the $s_y$ component oscillates only slightly.
In other words, the electron spin rotates in the $x-z$ plane.
After leaving the gate region, the electron possesses either the same  [Fig.~\ref{fig6}(a)]
or opposite spin [Fig.~\ref{fig6}(b)] to that of the injected electron.
In the main panel of Fig.~\ref{fig6}, we have plotted ratio $L_g/\lambda_{SO}$ of the gate
length $L_g$ to the precession length $\lambda_{SO}$ as a function of gate voltage.
The precession length $\lambda_{SO}$ has been calculated as follows:
$\lambda_{SO} = 2\pi/\Delta k$, where $\Delta k = |k^+-k^-|$.
The estimated values of the precession length are of the order of gate length,
in particular $\lambda_{SO} = 150$~nm for $V_g=1.55$~V
and 225~nm for $V_g=2.34$~V.
We note that in Ref.~\onlinecite{Fasth07} the value of the precession length $\lambda_{SO} \simeq 120$nm,
obtained  from the simple model\cite{Fasth07} used to elaborate the experimental data for the InAs nanowire,
is also comparable with the gate length ($L_g \simeq 100$~nm in Ref.~\onlinecite{Fasth07}).
Ratio $L_g/\lambda_{SO}$ is a linear function of the gate voltage,
parametrized as follows: $L_g/\lambda_{SO} = \alpha V_g$, where $\alpha = 0.65$~V$^{-1}$.
This means that $\lambda_{SO}$ is a decreasing function of the gate voltage
in agreement with the experimental data.\cite{Dhara09}
The linear dependence of ratio $L_g/\lambda_{SO}$ on $V_g$
allows us to determine the gate voltage that causes the given number
of spin rotations performed during the flow of the electron through the nanowire.
If $L_g/\lambda_{SO} = N$,  where $N = 1, 2, \ldots$, the electron spin performs the integer number
of rotations [Fig.~\ref{fig6}(a)] and the electron does not change its spin on the output.
If $L_g/\lambda_{SO} = N+1/2$, the number of electron spin rotations is half integer and
the electron changes the spin orientation from the initial (input) value $s_z =+\hbar/2$ to the final
(output) value $s_z= -\hbar/2$  [Fig.~\ref{fig6}(b)].
The number of spin rotations depends on the strength of the SOI that is controlled by the gate voltage.
\begin{figure}[ht]
\begin{center}
\includegraphics[scale=0.25, angle=0]{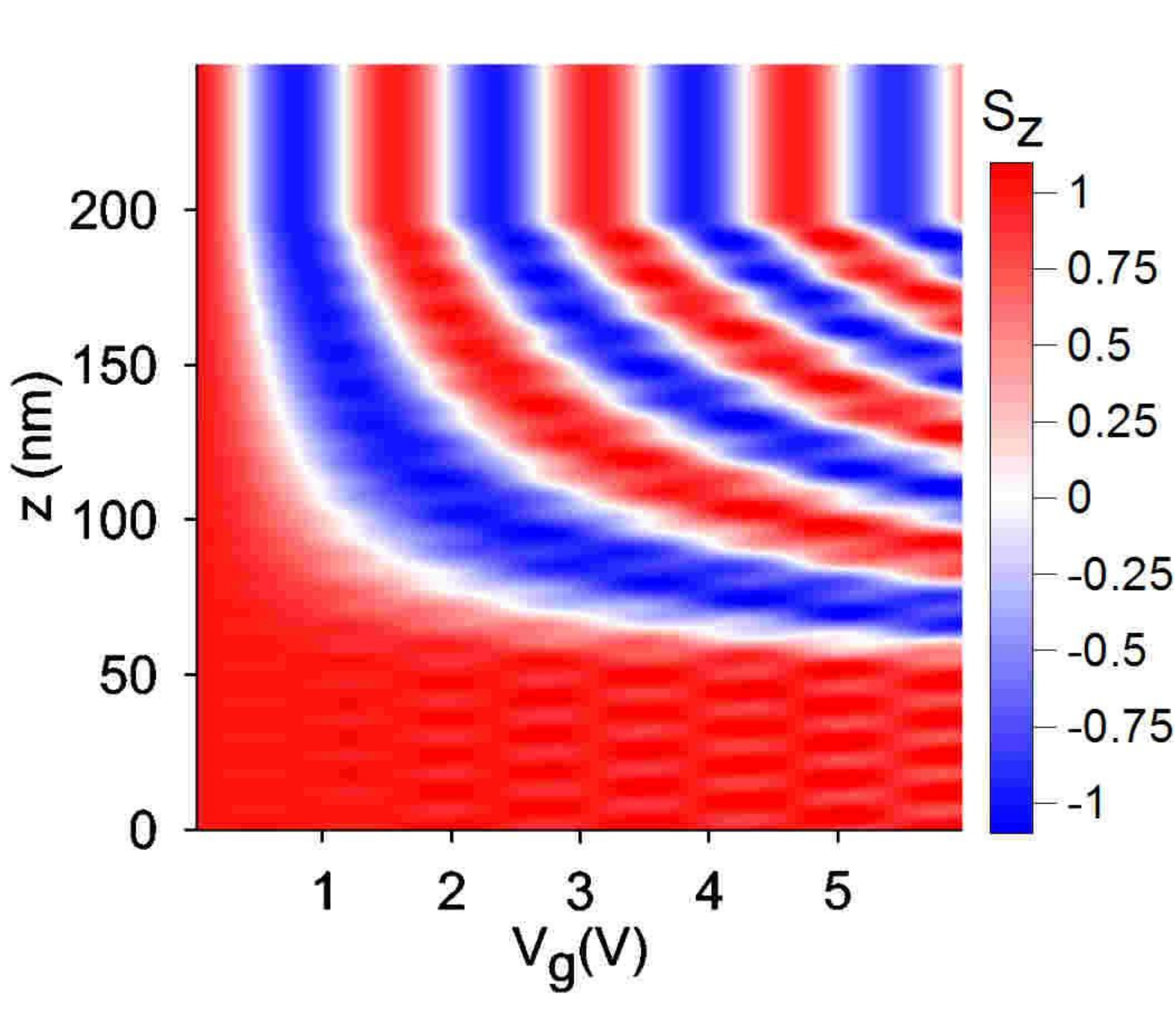}
\caption{Spin density $s_z$ as a function of the gate voltage $V_g$ and coordinate $z$.
The electrons with spin $s_z = +\hbar/2$ are injected  from the source at $z=0$ and the output spin states
of the electrons are detected in the drain at $z=250$~nm.}
\label{fig7}
\end{center}
\end{figure}

In order to determine the possibilities of the spin manipulation by the gate voltage
we have calculated the change of the spin density along the nanowire axis as a function of $V_g$
(Fig.~\ref{fig7}).  The electrons with spin $s_z=+\hbar/2$  injected from the source at $z=0$
are detected in the drain at $z=250$~nm.
Depending on the gate voltage the detected electrons can have different spins
that vary from $s_z =-\hbar/2$ through  $s_z=0$ to $s_z=+\hbar/2$.
Only for the suitably chosen gate voltage values the output electrons possess exactly the same
spin as the input electrons. For these gate-voltage values the spin transistor is in the on-state.
Figure \ref{fig7} also shows how to choose the gate voltage in order to obtain the output electrons with the opposite spin
to that of the injected electrons.
If the electrons change their spins in the nanowire, they cannot flow
through the device and the spin transistor is in the off-state.

\subsection{Comparison with experiment}

\begin{figure}[ht]
\begin{center}
\includegraphics[scale=0.3, angle=0]{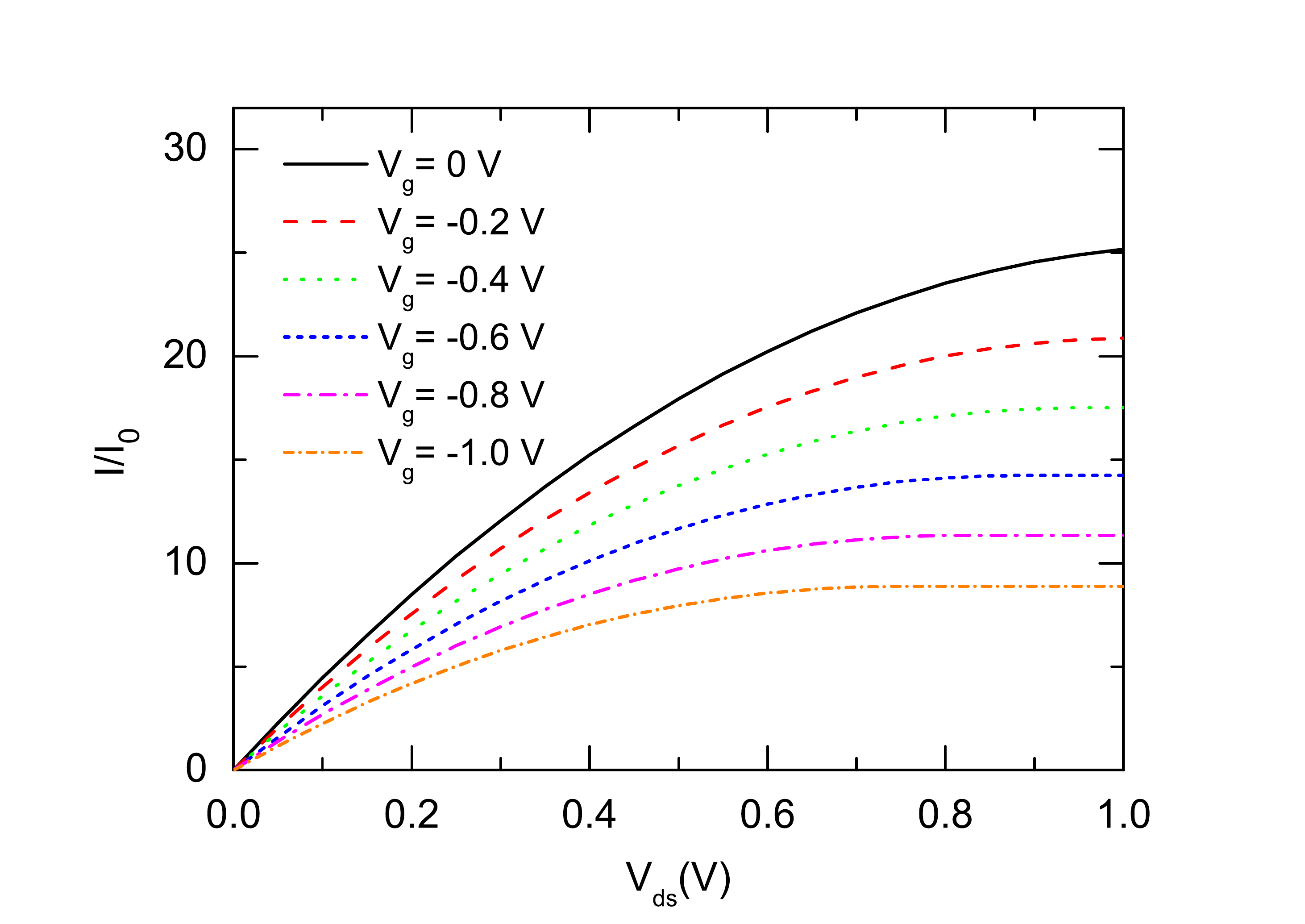}
\caption{Current $I$ as a function of drain-source voltage $V_{ds}$
and gate voltage $V_g$.  Polarization of the electrons in the contacts $P=0.4$,
$I_0 = 2e\mu/h$.}
\label{fig8}
\end{center}
\end{figure}

\begin{figure}[ht]
\begin{center}
\includegraphics[scale=0.3, angle=0]{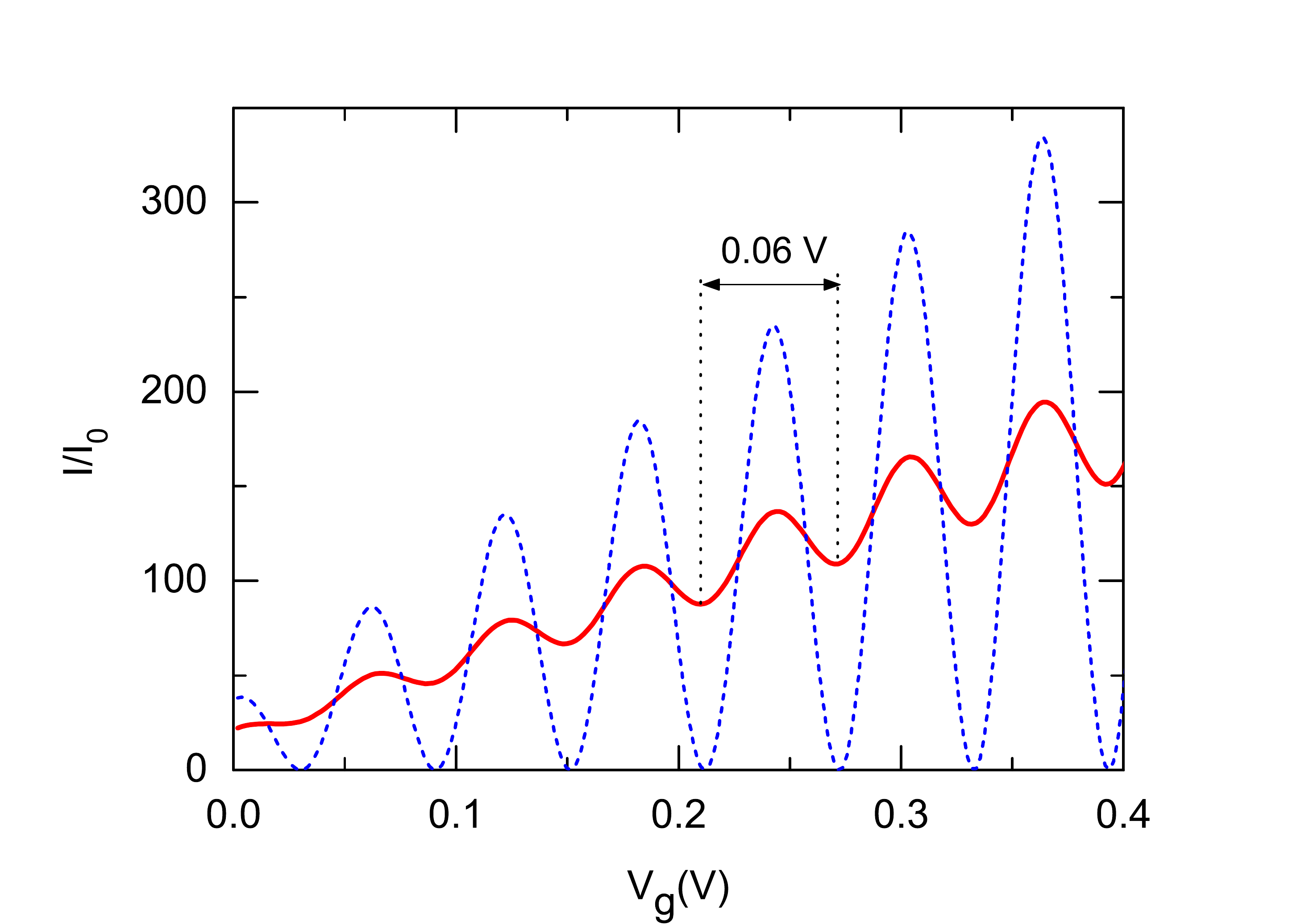}
\caption{Current $I$ as a function of gate voltage $V_g$ for $V_{ds}=0.6$~V
and for $P=0.4$ [solid (red) curve] and $P=1$ [dashed (blue) curve].
$I_0 = 2e\mu/h$.}
\label{fig9}
\end{center}
\end{figure}

In this subsection, we present the results of the calculations, which have been motivated
by the recent paper of Yoh et al.\cite{Yoh12} with the InAs nanowire spin transistor.
The operation of the spin transistor has been observed\cite{Yoh12} at the room temperature for the InAs nanowire
with length $L= 4$ $\mu$m and the gate with length $L_g = 2.8$ $\mu$m attached to the nanowire.
It is well known that for the nanowire with the length on the order of few $\mu$m the assumption of the ballistic transport
is not satisfied. Nevertheless, we have extended the ballistic transport model beyond the scope of its applicability in order to
simulate the spin transistor operation observed in the few-micrometer-long nanowire.
The agreement of the calculation results with the experiment can be treated
as an {\em a posteriori} justification of such extension.
The calculations have been carried out for the nanowire with the same geometry as that in
Ref.~\onlinecite{Yoh12} and for temperature $T=300$~K.
The influence of the temperature has been taken into account only by the Fermi-Dirac distribution
of the electrons in the contacts.   The electron-phonon scattering has been neglected.
In these calculations, we have applied the many-subband approximation,
where number $M$ of the subbands taken into account
was determined by the Fermi-Dirac distribution function [cf. Eq.~(\ref{I})],
and assumed the partial polarization of the electrons in the contacts
taking on $P=0.4$,
which corresponds to the electron spin polarization in Fe electrodes used in the experiment.\cite{Yoh12}
We have treated chemical potential $\mu$ as a fitting parameter and
adjusted its value ($\mu = 1.1$eV) in order to obtain the best agreement
between the calculated and measured current-voltage characteristics for $V_g =0$ (Fig.~\ref{fig8}).
The fitting of this single parameter has allowed us to
obtain the current-voltage characteristics for other values of the
gate voltage (Fig.~\ref{fig8}) and current vs gate voltage
oscillations (Fig.~\ref{fig9}) in a good agreement with the
experimental data.\cite{Yoh12}
The calculated current-voltage characteristics are plotted in Fig.~\ref{fig8}.
We see that the $I(V_{ds})$ curves (Fig.~\ref{fig8}) saturate at smaller drain-source voltage
if the gate voltage takes on the lower negative values.
The value of the saturated current decreases with the decreasing bias voltage.
This decrease results from the fact that the lowering of the gate
voltage inactivates the subsequent conduction channels, which leads to the drop of the current.

Figure \ref{fig9} displays the source-drain current as a function of the gate voltage
for constant drain-source voltage $V_{ds}=0.6$~V.
In Fig.~\ref{fig9}, we can distinguish the following two current components:
the first that monotonically increases with the gate voltage
and the second that oscillates as a function of the gate voltage.
The monotonically increasing current component results from the activation of  the subsequent conduction channels,
which occurs when we increase the gate voltage.
The oscillating current component results from the SOI induced precession of the electron spin in the vicinity of the gate.
The period of oscillations ($\Delta V_g = 0.06$~V) estimated from Fig.~\ref{fig9} agrees with the measured value.\cite{Yoh12}
The appearance of the oscillating current component can be explained as follows.
The total current flowing through the nanowire consists of the two spin-polarized currents
with the opposite spin polarization.  The contributions of these spin-polarized  currents to the total current are not equal
to each other, which results from the non-zero spin polarization of the electrons injected from the source
[cf. Eqs.~(\ref{I_up}) and (\ref{I_down})].
This means that the total current flowing through the nanowire is partially spin-polarized [cf. Eq.~(\ref{I_total})].
The SOI induced precession of the electron spin in the vicinity of the gate causes that
the following pairs of current components: ($I_{\uparrow \uparrow}$, $I_{\uparrow \downarrow}$) and
($I_{\downarrow \downarrow}$, $I_{\downarrow \uparrow}$)
oscillate in anti-phase as a function of the gate voltage.
These oscillations lead to the changes of the spin polarization of the current.
We note that the spin-polarized current can
flow through the drain without any reflection only
if the spin polarization of the current is the same as the spin polarization of the electrons in the drain.
Therefore, the total current reaches its maximal value for the gate voltage, for which this condition is satisfied.
Since the spin-polarized current oscillates as a function of the gate voltage,
we obtain the oscillations of the total current on the characteristics $I(V_g)$ (Fig.~\ref{fig9}).
The amplitude of these oscillations increases with the increasing spin polarization of the electrons in the contacts.
Figure~\ref{fig9} also shows that for the full spin polarization of the electrons in the contacts
the minimal values of the oscillating current are exactly equal to zero.
In Fig.~\ref{fig9}, as opposite to Fig.~\ref{fig5}(a), the amplitude of the current oscillations increases with the gate voltage,
which results from the activation of the subsequent conduction channels.

\section{Conclusions and Summary}

In the present paper, we have studied the two operation modes of the spin transistor:
(A) the ideal operation mode for the zero temperature with the full polarization of electron spins in the contacts
and (B) the more realistic operation mode for the room temperature with the partial polarization of electron spins in the contacts.
For both the operations modes we have found that the gate-voltage induced SOI leads to the oscillations
of the drain-source current as a function of gate voltage.
In case (A), the total current stops to flow for the appropriately adjusted gate voltage.
In case (B),  the total current  does not reach zero, but by the tuning of the gate
voltage we can obtain the rapid decrease of the total current to its minimal value, which is determined
by the spin polarization $P$ of the electrons in the contacts.
For both the modes we have determined the conditions for
the on/off states of the spin transistor in the gated nanowire.

We have also analyzed the spin precession induced by the SOI and demonstrated that this precession results from the superposition
of the rotations of $x$ and $z$ spin components.
The estimated precession length is of the order of the gate length and is inversely
proportional to the gate voltage.
Moreover, we have determined the effect of the spin polarization of electrons in  the contacts on
the operation of the spin transistor.

In summary, the theoretical description proposed in the present paper allows us to study
the physics underlying the operation of the spin transistor in the gated nanowire
with the Rashba spin-orbit interaction and provides the computational results
in a good agreement with the experimental data.

\end{document}